\documentclass[a4paper,mathleft]{revtex4}
\usepackage{graphicx}
\usepackage{times}
\usepackage{natbib}
\begin{document}

\title{An active fiber sensor for mirror vibration metrology 
in astronomical interferometers}

\author{S. Minardi$^{1,2}$\footnote{Corresponding author: stefano.minardi@uni-jena.de},  A. Chipouline$^{2}$, S. Kr\"amer$^{1,2}$, T. Pertsch$^{2}$ 
R. Follert$^{3}$, B. Stecklum$^{3}$,  R. Neuh\"auser$^{1}$
}

\affiliation{
1. Astrophysikalisches Institut und Universit\"ats-Sternwarte, Friedrich-Schiller-Universit\"at Jena, D-07745 Jena, Germany
\\
2. Institut f\"ur Angewandte Physik, Friedrich-Schiller-Universit\"at Jena, Max-Wien Platz 1, D-07743 Jena, Germany
\\ 
3. Th\"uringer Landessternwarte Tautenburg, Sternwarte 5, D-07778 Tautenburg, Germany}

\keywords{Astrointerferometry, astronomical instrumentation}

\begin{abstract}
We present a fiber sensor based on an active integrated component which could be effectively used to measure
the longitudinal vibration modes of telescope mirrors in an interferometric array. We demonstrate the possibility to measure vibrations with frequencies up to $\simeq 100$ Hz with a precision better than 10 nm.
\end{abstract}

\maketitle

\section{Introduction}
In the observation of faint targets, long-term fringe-tracking stability is perhaps the major challange which existing astronomical optical interferometers are currently facing. To insure such a stability, modern interferometric facilities implement active loops to compensate for the perturbations of the optical path difference (OPD) between the arms of the interferometer induced by the atmospheric turbulence and the vibrations of the optical elements of the interferometer. The power spectra
of these two noise contributions are usually well separated. In the near to mid-infrared band,
atmospheric turbulence contributes to the continuum part of the noise spectrum, mainly at
frequencies below 10-20 Hz. Such low frequency perturbations can be effectively
compensated by means of mechanical/piezoelectrical delay lines controlled by an error
signal generated from phase measurements on white-light interferograms of the target star
(Gai et al. 2004). On the contrary, mechanical vibrations of the interferometer
contribute typically to the discrete part of the noise spectrum with frequencies
ranging between 10 Hz and a few kHz (Sch\"oller 2007). These frequencies are too high to be compensated
by fringe tracking based on starlight, which usually requires integration times
of a few milliseconds to collect a significant signal. 
Several passive and active solutions have been proposed and applied to tackle this problem.
Among the passive methods, the removal of vibration sources, usage of vibration-damping optical mounts and of integrated photonic components for interferometric beam combination have demonstrated their effectiveness in improving the stability of astronomical interferometers (K\"ohler, L\'ev\^eque \& Gitton 2003;Coud\'e Du Foresto, Ridgway \& Mariotti 1997; Le Bouquin et al. 2004).
Despite these construction solutions, very high performance optical interferometry still requires active vibration compensation systems. This requirement is particularly critical in facilities using large collecting area telescopes such as the Very Large Telescope Interferometer (VLTI) or the Keck interferometer. Both facilities indeed impelement a vibration metrology system based on accelerometers placed on the most critical mirrors  (Colavita \& Wizinowich 2000; Hagenauer et al. 2008).  The information about the amplitude and frequency of the vibration mode is then forwarded to the fringe tracking system which counteract the estimated OPD perturbation with an appropriate piston applied to the delay line (Di Lieto et al. 2008).
The main limitation of this solution is that it can compensate only for a discrete set of periodic vibration modes.  
Laser metrology has also been used to stabilize the OPD perturbations occuring in the tunnels delivering the beams from the telescopes to the interferometric instruments, especially in settings involving dual star interferometry (Colavita et al. 1999; Colavita \& Wizinowich 2000; L\'ev\^eque 2000).
These settings however do not include the metrology of the mirrors of the telescopes, which are responsible for the largest amount of residuals of the OPD. 

In this paper, we propose and test a simple interferometric fiber sensor designed for laser metrology of vibrating optical elements in astronomical interferometers. The phase sensing is based on the active homodyne scheme (Jackson et al. 1980) and is implemented by means of fast electro-optical amplitude modulator. We report a precision in the OPD measurement better than 10 nm and a compensation bandwidth up to 100 Hz.

After a description of the experimental set-up (Section 2), we present stability tests of our fiber sensor (Section 3). The applications of the fiber sensor to astronomical interferometers are discussed in Section 4, and finally conclusions are drawn.

\section{Experimental setting}

\begin{figure}
\includegraphics[width=83mm]{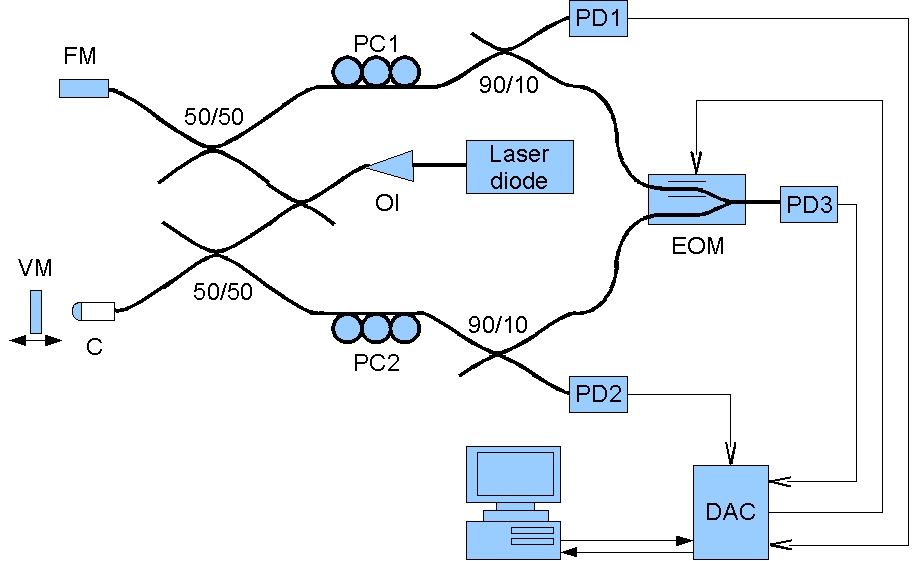}
\caption{\label{setup} Scheme of the tested fiber sensor. OI: optical insulator; FM: Faraday mirror; C: collimator; 
VM: vibrating mirror; PC: polarization controller; PD: photodiode; EOM: electrooptical modulator; 
DAC: digital-analogical converter}
\end{figure}

The general scheme of the prototype of the proposed interferometric fiber sensor is shown in Figure \ref{setup}. It consists basically in a reference arm, a metrology arm, a phase shifter and a beam combiner. The light from a stabilized semiconductor lasers source at $\lambda_0 = 1572$ nm is injected in the fibers by a two-level cascade of 50/50 fiber-splitters. Light from the reference arm is reflected back by a Faraday mirror. The metrology arm is terminated on a collimator and illuminates a mirror mounted on a loudspeaker. By driving the loudspeaker electrically we can simulate the vibrations of optical elements in a real astronomical interferometer. The orientation of the mirror is such that the reflected the light
is coupled back into the fiber.
Two 90/10-beamsplitters are used to couple out 10\% of light from each arm, just before the arms are recombined
to generate the interference signal. These pick-offs are used for the photometric correction of the interference data
(Hagenauer et al. 2000). The two high-troughput ends of the splitters are finally connected to the input of the electro-optical amplitude modulator
(model 92591 - JENOPTIK). This element can add an optical path difference between the two incoming channels and performs the beam combination required for the interferometric detection. A characterization of the performance of this element is found in the following paragraph. 
The contrast of the interferometric signal is maximized by adjusting the polarization controllers mounted across the fibers (PC)
and is monitored by a fast, low noise photodiode. The polarization alignment is required at the input of the modulators as we used standard SM28 fibers which are not polarization mantaining. Stable peak fringe visibilities as high as 95\% were achieved during
the experiments, indicating that polarization state drifts were not significant in our setting. The phase shift introduced by the vibrating mirror is measured by actively stabilizing the OPD in the fiber interferometer (active homodyne) (Jackson et al. 1980; Liu \& Measures 1992). This is achieved by feeding the interference signal to a digital PID controller which drives the amplitude modulator. The readout of the voltage applied to the modulator divided by the sensitivity of the phase modulator ($d\Phi/dV$) can provide an accurate measurement of the OPD.

\subsection{Characterization of the response of the elecro-optical modulator}

\begin{figure}
\includegraphics[width=83mm]{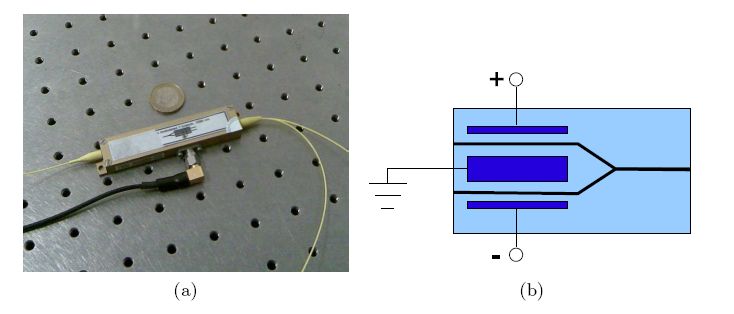}
\caption{\label{modulator}(a) The package of the amplitude modulator. (b) Scheme of the amplitude modulator.}
\end{figure}

A photograph and a simplified scheme of the amplitude modulator employed in our fiber sensor is shown in Figure \ref{modulator}.
The waveguiding and beam combination is accomplished by a Y-junction (Fig. \ref{modulator}.b) written by proton exchange
technique on a Lithium Niobate substrate. The waveguides are monomode in the range of wavelength 1300-1600 nm. 
A pair of electrodes is buried in the substrate next to one of the incoming waveguides. Due to the electrooptical
effect in Lithium Niobate, by applying a voltage across the electrodes it is possible to induce a refractive index variation in the
waveguides and therefore modify the optical path difference between the arms.
The overall phase diffrerence $\Phi$ at the wavelength $\lambda_0$ depends linearly on the applied voltage $V$ and can be written in the form:
\begin{equation}
\Phi=-\frac{\pi}{2\lambda_0}\frac{L}{d}\Gamma n_3^3 r_{33} V = - \frac{d\Phi}{d V}V
\end{equation}
where the refractive index along the cristallographic z--direction of the sample is $n_3$ and $r_{33}\simeq 33$ pm/V is the electrooptical coefficient. The distance between the electrodes is $d$ and $\Gamma$ is a coefficient which takes into account the inohomegeneites of the electrical field at the edge of the electrodes.  
The calibration of the sensitivity of the modulator is crucial for the determination of the phase shift introduced by the mirror vibrations. Its value was determined experimentally by recording the amplitude of the interference signal as a function of the amplitude of a 515 Hz periodic electrical signal applied to the phase modulator. At the working wavelength of 1572 nm we measured $d\Phi/dV=0.6196\pm 0.013$ rad/V. 

\begin{figure}
\includegraphics[width=83mm]{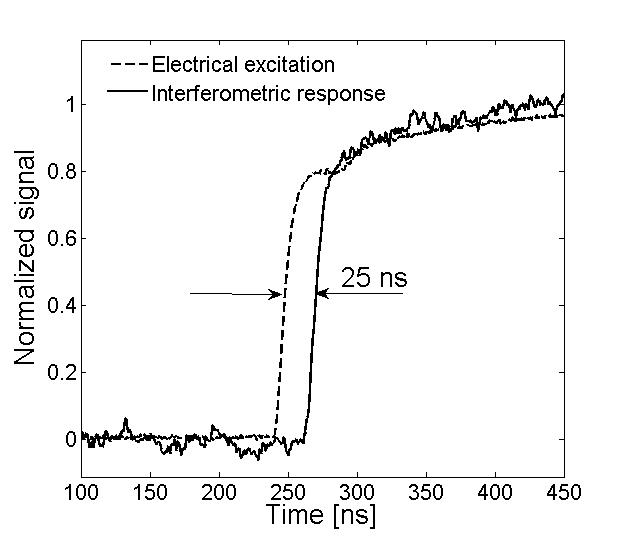}
\caption{\label{response} Time response of the modulator.}
\end{figure}

By modulating the applied voltage with a square function, we could also measure the actuation delay of the modulator, therefore assessing the potential for high frequency modulation. As shown in Fig. \ref{response}, the delay time between the voltage and the interference signal raising edges is about 25 ns. The actuation delay is therefore orders of magnitude faster than any existing mechanical delay line. 
The tradeoff is however a limited delay-stroke, due to the onset of electrical arching between electrodes at applied voltages larger than $\simeq 30$ V. For modulators of our type, the maximum OPD (equivalent in air) stroke declared by the manufacturer is $\pm 4.5 \mu$m.   
This stroke is however large enough to accomodate the maximum expected amplitude of the longitudinal vibration modes at the unit telescopes of VLTI which are below 4 $\mu$m (Richichi 2008). 

\subsection{PID controller}

The PID controller for the active stabilization of the interferometer is implemented on a LabView platform. An A/D converter (NI-PCI6251) operated at 68 kSamples/s per channel is used to manage the whole input/output of the controller (1 interferometric and 2 photometric inputs, 1 analogical output).
The incoming digitized signals are conditioned using a digital anti-alias filter with cut-off frequency set at 20 kHz. 
The cosinus of the phase difference between the arms is the control signal and is calculated from the interference signal and the calibrated photometric channels (Hagenauer et al. 2000). The set-point of the controller is 0, in order to stabilize the interferometer on the phase which allows for maximum linearity and sensitivity of the control loop. 
The correction signal for the modulator is then calculated using a standard forward integration algorithm. The optimal coefficients for the PID controller stabilization were determined empirically. The correction signal finally applied to the phase modulator after a suitable current amplification.

\section{Discussion of the results}

As mentioned before, in active homodyne the OPD between the arms of the interferometer is extracted from the feedback signal of the stabilized interferometer. For this reason, the performance of the phase measurement can be gauged from the parameters characterizing the stability of the fringe tracking.
In particular, the fluctuation of the stabilized interference signal can be related to the precision of the phase measurement while the bandwidth of the PID controller gives a quantitative assessment of the accuracy to which a modulated signal can be gauged with active homodyne.

\subsection{Stability of the vibration tracking}
\begin{figure}
\includegraphics[width=83mm]{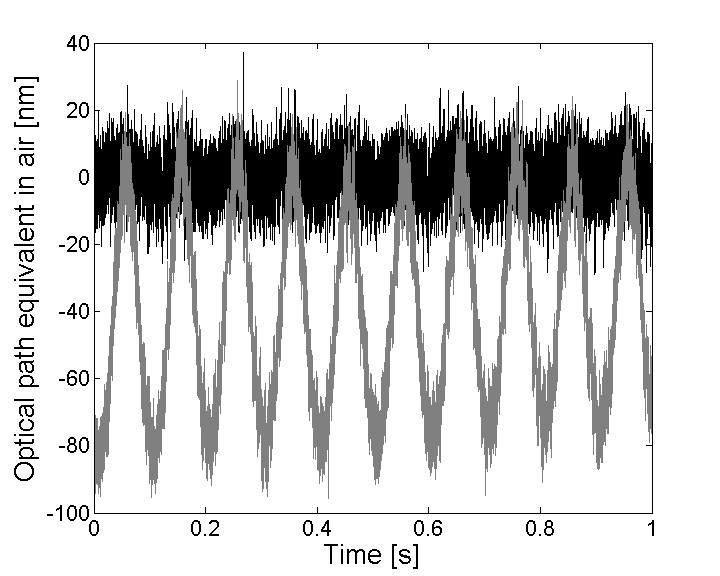}\\
\includegraphics[width=83mm]{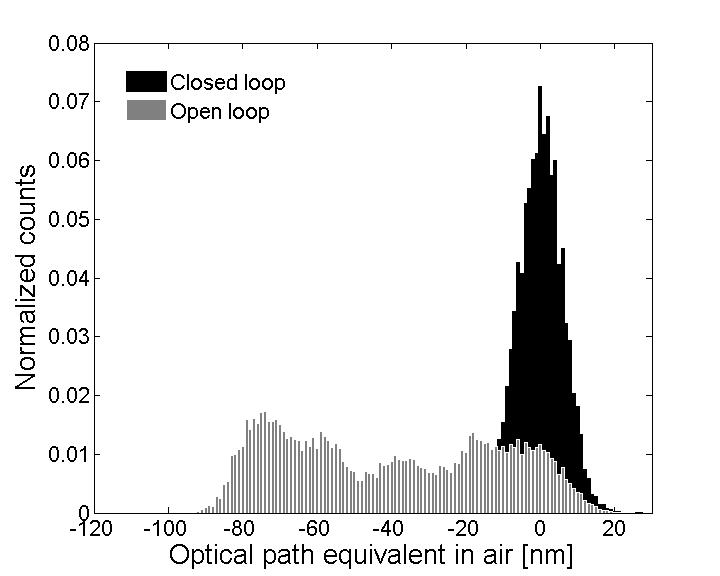}
\caption{\label{stability} The result of the stabilization of the fiber interferometer, displayed as time sequences of the interference signal (Top) and histogram (Bottom). Gray: open loop. Black: closed loop operation.}
\end{figure}

Figure 4 shows at glace the stabilization capabilities of our closed loop metrology system. A periodic modulation
of the OPD of 100 nm peak-to-peak was induced in the interferometer by means of the external mirror. The fringes could
be monitored in the open loop configuration (see Figure \ref{stability}(a)). By closing the loop, the residual r.m.s. fluctuation of the optical path dropped to 6 nm. A different view of the effect of the loop closing is provided in Fig. 4(b), where the
photodetector data over 1 s are displayed as histograms of the measured values.

The system could be stabilized over OPD fluctuations larger than $\lambda/2$, provided that the phase variation within the time response of the PID controller are smaller than $\simeq\pi/2$. This is because a single quadrature detection has monotonic response within a range of $-\pi/2$ to $+\pi/2$ from the stable work point. If the phase slips outside this range before the PID controller can compensate for it, the loop becomes unstable.  
The time constant of the PID controller was found to be $\tau\simeq 6$ ms (see next paragraph), thus setting the limit of operation of our active homodyne detector to signals with $(d\Phi/dt)_{\max} < 1.6\cdot10^3$ rad/s.

\subsection{Tracking bandwidth}

\begin{figure}
\includegraphics[width=83mm]{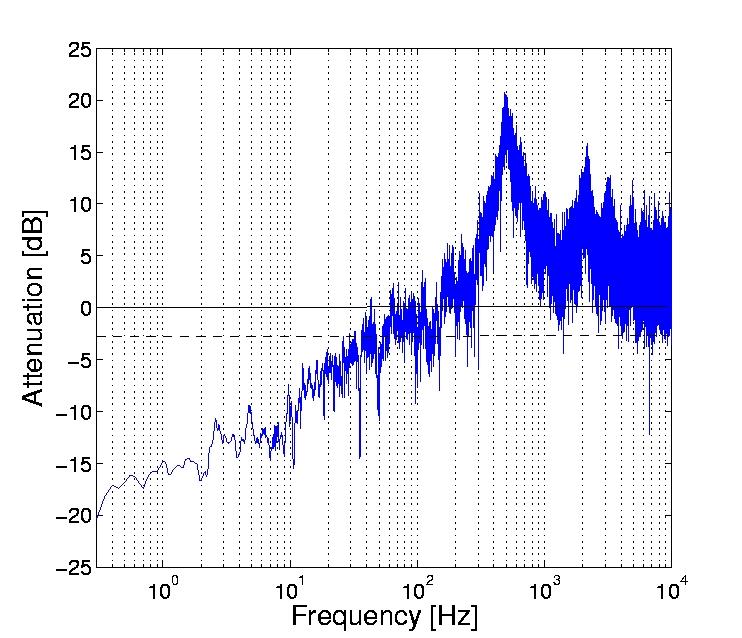}
\caption{\label{spectrum} Spectral response of the feedback loop.}
\end{figure}

The frequency response of the PID controller is characterized in term of the attenuation $A(\omega)$ of the signal which
we define as:
\begin{equation}
A(\omega) = 10 \log_{10} \left | \frac{a_{c}(\omega)}{a_{o}(\omega)}\right|
\end{equation}
where $a_{c}(\omega)$ and $a_{o}(\omega)$ represent the spectral amplitudes of the time traces of the photodiode
measured in closed and open loop configuration, respectively. We measured $A(\omega)$ experimentally from the Fourier
transform of two 10-second-long time traces of the natural variations of the OPD taken in open and closed loop configuration (Fig. \ref{spectrum}). The attenuation is high (-20 dB) for low frequencies and drops below -3 dB for exciting
frequencies $> 100$ Hz. 
The 0 gain point is reached at the cut-off frequency $f_0=160$ Hz. The inverse of this frequency is the time response of the PID controller $\tau=1/f_0\simeq 6$ ms. 
Numerical simulations of our control loop show that the time respose of the feedback loop is essentially limited by the total loop-delay $\tau_d$, the shortest achievable response time being $\simeq 10\tau_d$. In our case, measurements show that $\tau_d= 0.5$ ms. Because of the much shorter response of the phase modulator (see Fig.3), we conclude that the loop-delay is due essentially to the A/D conversion, communication and digital processing of the sampled interference data.

\section{Applications and limitations of the fiber sensor}

The main advantage of the design of our vibration sensor is the implementation with optical fibers which both provides flexibility and miniaturization.
Fibers can be easyly delivered to telescopes and the vibretion metrology installed without affecting significantly the collection capability of the telescope. Moreover, the wavelength for the metrology can be chosen within the atmospheric absorption bands so that scattered laser light could be easily rejected in the astronomical instrument by using the standard series of bandpass filters.

\begin{figure}
\includegraphics[width=83mm]{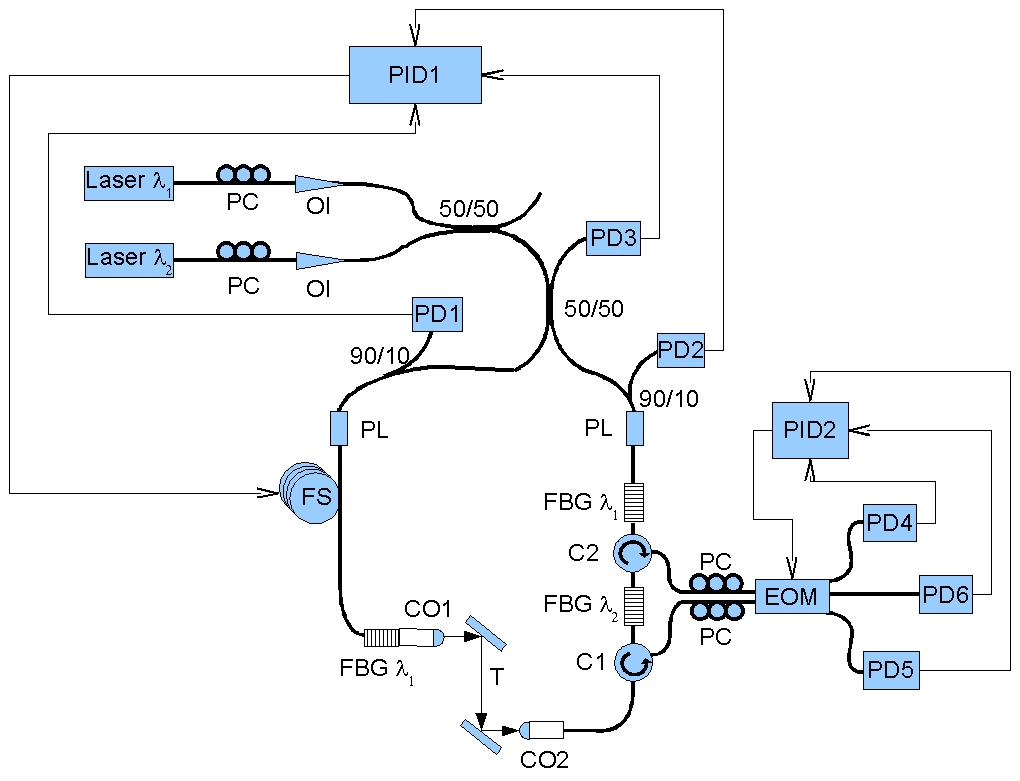}
\caption{ \label{updatedscheme} A possible scheme for a real metrology system for telescopes in an interferometer. C: circulator, CO: collimator, EOM: electro-optical modulator, FBG: Fiber Bragg Grating, FS: fiber stretcher, PC: polarization controller, PD: Photodiodes, PID: proportional-integral-differentical controller, OI: optical insulator, PL: in-line polarizer, T: optical train of the telescope.}
\end{figure}
We stress however that our simple demonstrator cannot distinguish between drifts of the OPD due to temperature or stress variations in the fiber and those originating from the moving mirror. Thermal/mechanical OPD drifts can be kept to negligible levels in small fiber interferometers (arm length $\simeq 10$m) in controlled environments such as a laboratory. However, significant OPD drifts of several millimeters are expected in large scale fiber interferometers (arm length: $\simeq 100$ m) deployed in non controlled environments (Kotani et al. 2005). A more complex scheme implementing srtabilization with fiber stretchers and an internal metrology of the interfereometer (Delage, Reynaud \& Lannes 2006; Lin et al. 2004) should be considered in this case.
A possible scheme for a realistic implementation of the vibration sensor is depicted in Fig. 6.
Two lasers emitting at the wavelengths of $\lambda_1$ and $\lambda_2$ are used in this scheme. The laser channel at $\lambda_1$ is reflected by fiber Bragg gratings (FBG) just before the collimators and used to measure the OPD between the two collimators. The homodyne phase measurement is accomplished by the photodiodes 1 to 3 (2 photometric and 1 interference channel). 
The homodyne phase signal is used to drive a feedback loop controlling a fiber stretcher. Electro-optical modulators cannot be used in this case, due to their limited OPD stroke. 

The light emitted by the laser at wavelength $\lambda_2$ propagates throught the fiber Bragg grating in arm 1 and is out-coupled to the telescope by means of a collimator. After reflecting on the mirrors of the telescope, the probe beam is coupled into arm 2 of the fiber interferometer by a second collimator and is combined with the reference channel in the electrooptical modulator. The incoming light from arm 1 and its reference propagating in arm 2 are diverted from arm 2 to the EOM by means of a combination of a FBG at $\lambda_2$ and two circulators. Photodiodes 4 to 6 will provide also in this case a homodyne signal needed for the stabilization of the EOM, and thus the phase measurement of the vibrations of the mirrors. 

As mentioned before, our prototype is limited in frequency response by the loop-delay introduced by our PID controller. Higher bandwidths (and thus a more complete coverage of the 100-1000 Hz range) can be easily obtained by reducing this loop-delay. Faster AD converters or analogical PID units can be used to extend the frequency coverage of the vibration detector. 

\section{Conclusion} 

We have demonstarted the feasibility of a mirror vibration fiber sensor based on active homodyne. The instrument can be deployed for vibration measurements in astronomical interferometers, thus providing useful data for their active stabilization.
Our experiments show that the scheme allows the determination of the optical path with a precision of at least 10 nm over a range of a few microns. 
We point out that our instrument combined with a feedback loop acting on the delay line of an astro-interferometer like VLTI could greatly improve the visibility of the fringes in the fringe tracker. 
This in turn will improve the fringe stability of the instrument, allowing for high dynamical range observations, e.g. detection of very faint objects very close to bright stars (planets in orbit around other
stars) or of surface structure on other stars (signature of convection) (Neuh\"auser et al. 2008).
Finally, we remark that for the implementation of the sensor, only off-the-shelf integrated optical components have been used. This makes our design economically more advantageous when compared to metrology solutions based on bulk optics and/or custom designed integrated optical components. 

SM acknowledges support from the EU in the FP6 MC ToK project MTKD-CT-2006-042514.

\bibliographystyle{agsm}

\newpage

\end{document}